\documentclass[aps,prb,twocolumn,superscriptaddress,showpacs]{revtex4-2}

\usepackage{amsmath,amssymb,wasysym,graphicx}
\usepackage[utf8]{inputenc}
\usepackage[T1]{fontenc}
\usepackage{xcolor}

\IfFileExists{newtxtext.sty}
{\usepackage{newtxtext,newtxmath}}
{\IfFileExists{stix.sty}
	{\usepackage{stix}}
	{\IfFileExists{mathptmx.sty}
		{\usepackage{mathptmx}}{} } }

\usepackage{textcomp}

\usepackage{bm}

\IfFileExists{siunitx.sty}{\usepackage{booktabs,siunitx}}{}

\usepackage{color}
\definecolor{LinkColor}{rgb}{0.256,0.439,0.588}
\usepackage{hyperref}
\hypersetup{
	pdfauthor={good guys},
	pdftitle={good title},
	colorlinks=true,
	citecolor=LinkColor,
	linkcolor=LinkColor,
	urlcolor=LinkColor
}

\newcommand{\beq} {\begin{equation}}
\newcommand{\eeq} {\end{equation}}
\newcommand{\bea} {\begin{eqnarray}}
\newcommand{\eea} {\end{eqnarray}}
\newcommand{\be} {\begin{equation}}
\newcommand{\ee} {\end{equation}}

\newcommand{\ket}[1]{\left|#1\right>}
\newcommand{\bra}[1]{\left<#1\right|}

\begin{document}
\title{Quantum entanglement of fermionic symmetry-enriched quantum critical points in one dimension}

\author{Wen-Hao Zhong}
\affiliation{Department of Physics, Fuzhou University, Fuzhou 350116, Fujian, China}
\affiliation{Institute for Advanced Study in Physics and School of Physics, Zhejiang University, Hangzhou 310058, China}

\author{Hai-Qing Lin}
\affiliation{Institute for Advanced Study in Physics and School of Physics, Zhejiang University, Hangzhou 310058, China}

\author{Xue-Jia Yu}
\email{xuejiayu@fzu.edu.cn}
\affiliation{Department of Physics, Fuzhou University, Fuzhou 350116, Fujian, China}
\affiliation{Fujian Key Laboratory of Quantum Information and Quantum Optics,
College of Physics and Information Engineering,
Fuzhou University, Fuzhou, Fujian 350108, China}

\date{\today}

\begin{abstract}
Quantum entanglement can be an effective diagnostic tool for probing topological phases protected by global symmetries. Recently, the notion of nontrivial topology in critical systems has been proposed and is attracting growing attention. In this work, as a concrete example, we explore the quantum entanglement properties of fermionic symmetry-enriched quantum critical points by constructing exactly solvable models based on stacked multiple Kitaev chains. We first analytically establish the global phase diagram using entanglement entropy and reveal three topologically distinct gapped phases with different winding numbers, along with three topologically distinct transition lines separating them. Importantly, we unambiguously demonstrate that two transition lines exhibit fundamentally different topological properties despite sharing the same central charge. Specifically, they display nontrivial topological degeneracy in the entanglement spectrum under periodic boundary conditions, thereby generalizing the Li-Haldane bulk-boundary correspondence to a broader class of fermionic symmetry-enriched criticality. Additionally, we identify a novel Lifshitz multicritical point at the intersection of the three transition lines, which also exhibits nontrivial topological degeneracy. This work provides a valuable reference for investigating gapless topological phases of matter from the perspective of quantum entanglement.  

\end{abstract}

\maketitle

\section{INTRODUCTION}
\label{sec:introduction}
Topological phases of matter form a cornerstone in modern condensed matter physics and quantum information science, extending beyond the traditional Landau-Ginzburg symmetry-breaking paradigm~\cite{Qi2011rmp,Hasan2010rmp,Wen2017rmp,senthil2015symmetry}. To date, most well-understood topological phases have focused on gapped quantum systems, ranging from symmetry-protected topological (SPT) phases, which include both free fermions~\cite{Kane2005prl_a,Kane2005prl_b,Bernevig2006Science,Markus2007Science} and strongly correlated many-body systems~\cite{Gu2009prb,Chen2013prb,Chen2012Science,Pollmann2012prb}, to intrinsically topologically ordered phases features by fractionalized anyonic excitations~\cite{Wen1990TO,wen1995topological,Zhou2017rmp,Savary_2017,Wen2019Science,Broholm2020Science}. However, recent advances suggest that symmetry-protected topology can persist even in gapless systems, including continuous quantum critical points (QCPs) and stable critical phases, collectively referred to as gapless symmetry-protected topological (gSPT) phases~\cite{cheng2011prb,fidkowski2011prb,kestner2011prb,keselman2015prb,ruhman2017prb,parker2018prb,JIANG2018753,keselman2018prb,scaffidi2017prx,thorngren2021prb,verresen2021prx,verresen2020topologyedgestatessurvive,DuquePRB2021,yu2022prl,yu2024prl,parker2019prl,yu2024prb,Yang2025CP,zhong2024pra,umberto2021sci_post,friedman2022prb,li2023intrinsicallypurelygaplesssptnoninvertibleduality,huang2023topologicalholographyquantumcriticality,wen2023prb,Milad2022prb_a,Milad2022prb_b,wen2023classification11dgaplesssymmetry,wen2024stringcondensationtopologicalholography,li2024sci_post,huang2024fermionicquantumcriticalitylens,su2024prb,zhang2024pra,ando2024gaugetheorymixedstate,zhou2024floquetenrichednontrivialtopologyquantum,li2024noninvertiblesymmetryenrichedquantumcritical,yu2025gaplesssymmetryprotectedtopologicalstates,tan2025exploringnontrivialtopologyquantum,yang2025deconfinedcriticalityintrinsicallygapless}. Among these, nontrivial topology in free fermion critical systems have attracted particular attention due to their exact solvability~\cite{verresen2018prl,verresen2020topologyedgestatessurvive,zhou2024floquetenrichednontrivialtopologyquantum}, making them valuable for establishing a general theory of fermionic symmetry-enriched QCPs in arbitrary dimensions. In contrast, higher-dimensional interacting critical systems remain challenging due to the lack of efficient analytical and numerical tools.

A central task is to classify and characterize topological phases in a unified framework. Traditionally, gapped topological phases are considered to share the same bulk properties as trivial quantum paramagnets, exhibiting topological distinctions only when open boundaries are present~\cite{Gu2009prb,Chen2012Science,Chen2013prb,Wen2017rmp}. Similarly, in the case of symmetry-enriched QCPs, the bulk universality class of topological gapless systems is identical to their trivial counterparts, while symmetry-protected edge modes emerge under open boundary conditions~\cite{verresen2021prx,yu2022prl}. As a result, these topological phases cannot be detected using local observables in the bulk. However, previous studies~\cite{LAFLORENCIE20161,Li2008prl,Pollmann2012prb,Turner2011prb,Pollmann2010prb,Isakov2011NP,Qi2012prl,Hsieh2014prl,Cho_2017,Chandran2014prl,Yao2010prl,Joshi2023Nature,zache2022entanglement,Kokail2021NP,yu2024prl} suggest that quantum entanglement serves as a definitive signature of both topological and critical phases, with entanglement entropy and the entanglement spectrum being the most representative physical quantities. The former captures universal data, such as the central charge, in the conformal critical points within certain universality classes~\cite{francesco2012conformal,ginsparg1988appliedconformalfieldtheory}, while the latter encodes nontrivial topological degeneracy and acts as a fingerprint for identifying topological phases~\cite{Li2008prl,Pollmann2010prb,yu2024prl}. Specifically, the entanglement spectrum under periodic boundary conditions, known as the bulk entanglement spectrum, encodes the topological degeneracy of the energy spectrum under open boundary conditions, which is referred to as the Li-Haldane conjecture~\cite{Li2008prl}, has recently been generalized to bosonic symmetry-enriched QCPs in quantum spin chains~\cite{yu2024prl,zhang2024pra}.  Therefore, the bulk entanglement spectrum provides an effective diagnostic probe that does not require boundaries and preserves all symmetries of the system, including lattice symmetries.  

A promising direction is to use the information-theoretic concept of entanglement to characterize free fermionic symmetry-enriched QCPs, which allows us to be analytically tractable and avoids the computational challenges associated with gapless many-body systems. Consequently, similar to the extensively studied quantum entanglement properties of free-fermion topological insulators and superconductors~\cite{Turner2011prb,Qi2012prl,Fidkowski2010prl,Hsieh2014prl,Cho_2017}, we can explore analogous questions for free-fermion symmetry-enriched critical states: What are the quantum entanglement properties of these states? More importantly, can we uncover novel entanglement signatures at the transition points between these fermionic symmetry-enriched QCPs?

In this work, we address the series of questions outlined above using an exactly solvable Majorana fermion model, constructed by stacking multiple Kitaev chains. Using entanglement entropy as a diagnostic, we analytically establish the global phase diagram, which exhibits three types of gapped phases characterized by distinct topological winding numbers ($\omega = 0,1,2$). The three conformal critical lines, which separate the topologically distinct gapped phases, can be determined analytically, while a nonconformal Lifshitz multicritical point emerges at the intersection of these transition lines. Through finite-size scaling analysis of entanglement entropy, we find that both transitions between the $\omega = 1$ topological superconducting (TSC) phase and either the trivial or the $\omega = 2$ TSC phase belong to the Majorana universality class with central charge $c = 1/2$. In contrast, the transition between the trivial and the $\omega = 2$ TSC phase belongs to the Dirac universality class with $c = 1$. Furthermore, we employ the bulk entanglement spectrum as a probe to characterize the nontrivial topology of these quantum critical lines. Our numerical results unambiguously demonstrate that only the transition lines between TSC phases with different winding numbers exhibit topological degeneracy in the entanglement spectrum, corresponding to fermionic symmetry-enriched QCP. In contrast, the remaining two transition lines are both topologically trivial, even though one of them shares the same central charge ($c = 1/2$) as the topologically nontrivial one. Finally, we explore the topological properties at the Lifshitz multicritical point by examining its bulk entanglement spectrum, revealing a topological Lifshitz multicritical point with nontrivial degeneracy.  

The paper is organized as follows: Section.~\ref{sec:model} introduces the physics of the Majorana $\alpha$ chain and presents the exactly solvable lattice model we focus on. Section.~\ref{sec:phase} (a) provides a self-contained review of entanglement entropy and spectrum, along with the calculation scheme for free fermion systems. Section.~\ref{sec:phase} (b) maps out the global phase diagram of the model using entanglement entropy. As a warm-up, we first examine the entanglement spectrum to identify different gapped topological phases in Section.~\ref{sec:phase} (c), and characterize critical and topological properties at three transition lines and the multicritical point in Section.~\ref{sec:phase} (d). The conclusions are presented in Section.~\ref{sec:con}, and additional data from our analytical and numerical calculations are provided in the Appendix.  

\section{MODEL: MAJORANA $\alpha$ CHAIN}
\label{sec:model}
\begin{figure}
    \centering
    \includegraphics[width=0.8\linewidth]{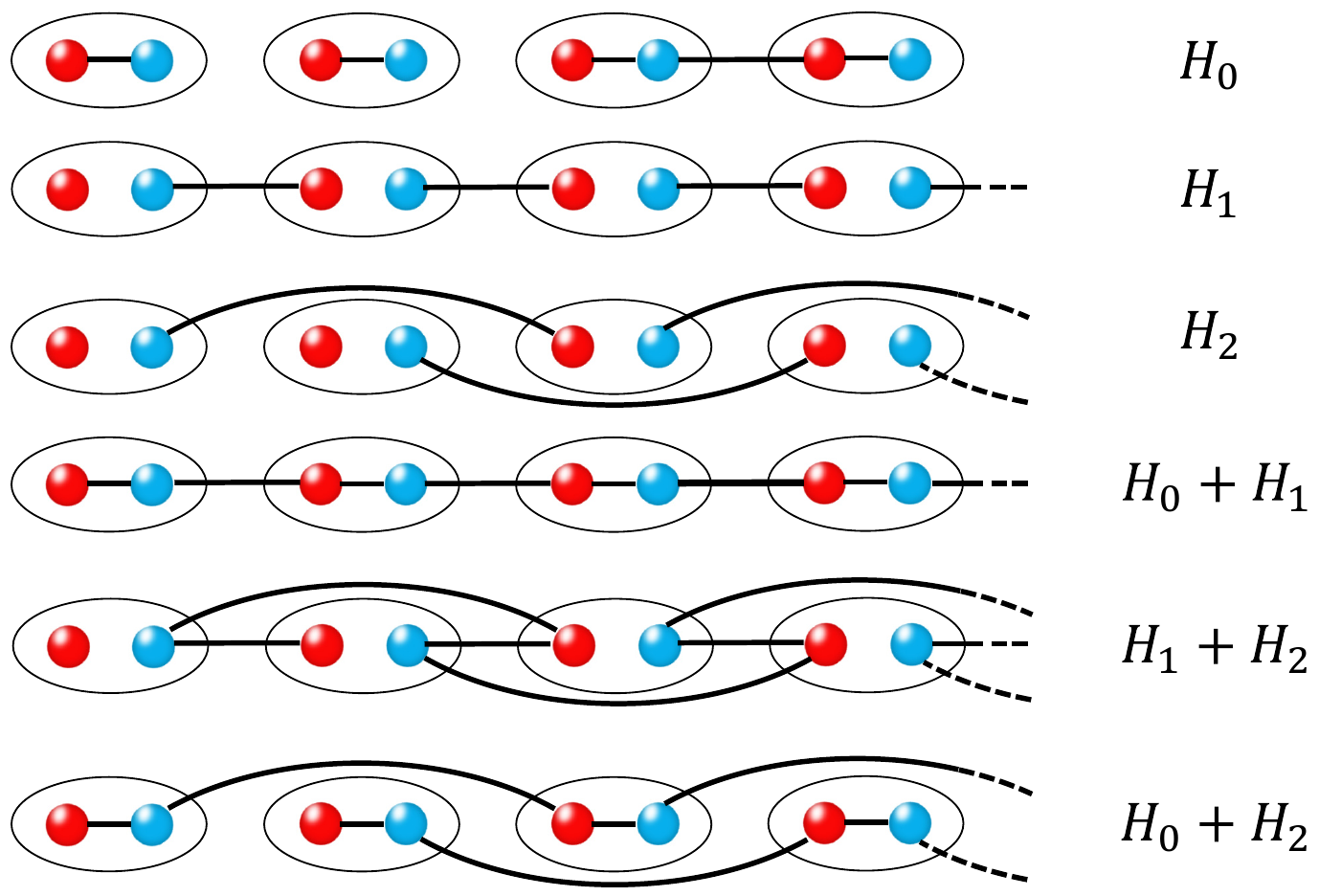}
    \caption{(Color Online) The representative Hamiltonian of the Majorana $\alpha$ chain for both gapped and gapless phases is shown. Fermionic chains with $ L = 4 $ unit cells are depicted, where each unit cell consists of two sublattice sites, represented by red and blue circles. The unpaired Majorana fermions at the edges illustrate the topological nature of both the gapped and gapless states.}
    \label{fig1}
\end{figure}

The simplest model that exhibits both gapped SPT and symmetry-enriched QCPs is constructed by stacking $\alpha$ Kitaev chains on top of each other, which is equivalent to coupling Majorana modes over a distance $\alpha \in \mathbb{Z}$. This model, known as the Majorana $\alpha$ chain (abbreviated as $\alpha$ chain)~\cite{verresen2018prl,Verresen2017prb,Choi_2024}, is described by the Hamiltonian:  
\begin{equation}
\label{E1}
H_{\alpha} = i \sum_{n} \tilde{\gamma}_{n} \gamma_{n+\alpha}, \quad \alpha \in \mathbb{Z},
\end{equation}  
where the Majorana operators are expressed in terms of complex fermion operators $c$ as $\gamma_n = c_{n}^{\dagger} + c_{n}$ and $\tilde{\gamma}_{n} = i (c^{\dagger}_{n} - c_{n})$. 
Each unit cell consists of two Majorana fermions, $\gamma_n$ (red) and $\tilde{\gamma}_n$ (blue), as illustrated in Fig.~\ref{fig1}. The $\alpha$-chain preserves both particle-hole $\mathcal{C}$ and time-reversal symmetry $\mathcal{T}$, which satisfy $\mathcal{T}^{2} = \mathcal{C}^{2} = +1$. Consequently, the model belongs to the BDI symmetry class~\cite{Altland1997prb}. As a result, the topological nature of $\alpha$ chains can be characterized by a discrete topological winding number $\omega = 0, 1, 2, \dots$. Moreover, linear combinations of different $\alpha$ chains can give rise to topologically distinct Majorana conformal field theory (CFT), leading to the concept of fermionic symmetry-enriched QCPs~\cite{verresen2018prl,yu2024prb}. Additionally, a nonconformal tricritical point emerges at the transition between these topologically distinct conformal critical lines.

Intuitively, the terms in $ H_{\alpha} $ serve as fermionic duals of Ising spin Hamiltonians. For example, $ H_0 $ corresponds to the transverse field $ \sigma^x_n $, representing a trivial band insulator. $ H_1 $ describes the Ising spin-spin interaction $ \sigma^{z}_{n} \sigma^{z}_{n+1} $, while $ H_2 $ represents a cluster interaction $ \sigma^{z}_{n} \sigma^{x}_{n+1} \sigma^{z}_{n+2} $, whose ground state belongs to a gapped cluster SPT phase~\cite{Son_2011,Smacchia2011pra,Guo2022pra,Li2025pra}.  

Moreover, the composite Hamiltonians $ H_0 + H_1 $ and $ H_1 + H_2 $ correspond to topologically distinct critical Majorana chains, which are the fermionic duals of the critical transverse and cluster Ising models. These two critical points are both characterized by the same Ising CFT but exhibit distinct topological properties, such as degenerate edge modes even at criticality. This phenomenon is referred to as symmetry-enriched quantum criticality~\cite{verresen2021prx}. Under the Jordan-Wigner duality, these nontrivial topological critical points can also emerge in free fermion systems, giving rise to the fermionic symmetry-enriched QCPs.

We would like to emphasize that several important subtleties remain between symmetry-enriched quantum critical points in free fermion systems and those in interacting spin chains, which are elaborated upon below: i) The emergence of topological edge states in critical free fermion systems arises from a novel mechanism known as kinetic inversion, which is fundamentally distinct from the symmetry-enriched CFT framework used for interacting spin chains. Crucially, this mechanism does \emph{not} rely on conformal symmetry, which is essential in the symmetry-enriched CFT description. ii) In interacting critical spin chains, the spatial decay of topological edge states can exhibit algebraic localization—an exotic feature that does not occur in free fermionic systems, where edge states remain exponentially localized. iii) The exactly solvable nature of free fermion systems provides a valuable avenue for exploring gapless topology in higher dimensions, offering a platform that is tractable both analytically and numerically. More importantly, for free fermionic tight-binding models, the winding number—which is widely used to classify gapped topological insulators—is ill-defined in the gapless regime and cannot be directly generalized. As a result, the classification of gapless topological phases in free fermion systems remains an open question, particularly in higher dimensions, despite recent progress~\cite{verresen2018prl,verresen2020topologyedgestatessurvive,verresen2021prx}. In contrast, the entanglement spectrum—the central diagnostic used in our work—can unambiguously distinguish between topologically trivial and nontrivial critical points, both in free fermion systems and in interacting critical spin chains~\cite{yu2024prl}. Therefore, it serves as a robust and unique fingerprint for classifying topologically nontrivial critical points in gapless free fermion systems for arbitrary dimensions. A systematic exploration of the entanglement spectrum in gapless topological states of free fermion systems in general dimensions will be presented in our upcoming work.

In this work, to systematically investigate the entanglement properties of fermionic symmetry-enriched QCPs, we focus on the simplest exactly solvable Hamiltonian that incorporates a linear combination of $ \alpha $-chains, given by $H = g_0 H_0 - g_1 H_1 + g_2 H_2$, where, without loss of generality, we set $ g_0 = 1.0 $ as the energy unit.  To understand the underlying quantum phases of this model, let us first consider the limiting cases:  1) When $ H_0 $ dominates, the ground state is a trivial product state. 2) When $ H_1 $ dominates, the system realizes a topological superconducting phase with a winding number $ \omega = 1 $, which is the fermionic dual of an Ising spontaneous symmetry breaking (SSB) phase with two-fold ground-state degeneracy. 3) When $H_2 $ dominates, the system remains in a topological superconducting phase but with a winding number $\omega = 2 $. The competition among $ H_0 $, $ H_1 $, and $ H_2 $ gives rise to a rich global phase diagram with intricate quantum entanglement properties, which are the central focus of this work.

\section{RESULTS}
\label{sec:phase}

\subsection{Quantum entanglement of free fermion systems}
To explore the quantum entanglement properties of fermionic symmetry-enriched QCPs, we first introduce the fundamental concepts of entanglement entropy and entanglement spectrum and then provide numerical details on how to compute these quantities in free fermion systems.

In quantum many-body systems, entanglement entropy serves as a key indicator of phase transitions by extracting relevant information from the ground-state wavefunction $ \ket{\varphi_{0}} $. \color{black} Typically, it is defined by partitioning the system into two subsystems, A and B, and computing the reduced density matrix for subsystem A by tracing out the degrees of freedom of subsystem B:  
\begin{equation}
\label{E7}
\rho_{\rm{A}} = {\rm{Tr}}_{\rm{B}}(\ket{\varphi_{0}}\bra{\varphi_{0}}).
\end{equation}
The entanglement entropy, measuring the entanglement between parts A and B, is then expressed as:
\begin{equation}
\label{E8}
S_{\rm{A}} = -{\rm{Tr}}\left[\rho_{\rm{A}}{\rm{ln}}(\rho_{\rm{A}})\right].
\end{equation}
which is evaluated in terms of the eigenvalues of $\rho_{\rm{A}}$. For a one-dimensional local quantum system with periodic boundary conditions, CFT suggests that the entanglement entropy for subsystem A with size $l$ follows the finite-size scaling behavior~\cite{francesco2012conformal,ginsparg1988appliedconformalfieldtheory}
\begin{equation}
\label{E9}
S_{l} \sim \frac{c}{3}\text{ln}\left[\frac{N}{\pi}{\rm{sin}}\left(\frac{\pi l}{N}\right)\right] + S^{\prime}.
\end{equation}
where $c$ is the central charge, which varies for different universality classes, and $S^{\prime}$ is a non-universal constant. For half chain entanglement entropy $S_{l=N/2}$, the finite-size scaling relation reduced to $S(N) \sim \frac{c}{3} \text{ln} N + S_0$. Another important entanglement-related quantity, known as the entanglement spectrum, consists of the eigenvalues of the modular or entanglement Hamiltonian $H_{\rm{A}}$, which is related to the reduced density matrix $\rho_{\rm{A}}$ of the subsystem A by $\rho_{\rm{A}} = e^{-H_{\rm{A}}}$.These two fundamental entanglement measures—entanglement entropy and entanglement spectrum—serve as powerful tools for diagnosing both critical and topological properties in quantum many-body systems.

Calculating these two entanglement quantities is generally challenging. However, in free fermion systems, entanglement-based quantities can be computed analytically using the correlation matrix technique, which we briefly review below. For the ground state $|\Phi \rangle$ of free fermion system with half-filling, the total density matrix can be expressed in terms of correlation matrix~\cite{Vidal2003prl,Peschel_2009,Zhou2022prr,peschel2003JOPA,cheong2003,chang2020prr,ye2016prb}:
\begin{equation}
    \label{E2}
    \rho = |\Phi \rangle \langle \Phi | = \det(I-\mathcal{G})\exp\left(\sum_{ij}\left[\ln\mathcal{G}(I-\mathcal{G})^{-1}\right]_{ij} \Psi_i^\dagger \Psi_j \right),
\end{equation}
where $\{\Psi_i\}$ is the Nambu basis which defined as $\Psi = (c_1, c_2, \dots, c_N, c_1^{\dagger}, c_2^{\dagger}, \dots, c_N^\dagger)^T$, and $\mathcal{G}$ is the correlation matrix, given by
\begin{equation}
    \label{E3}
        \mathcal{G}_{ij} =  
    \begin{pmatrix} 
        \langle c_i^\dagger c_j \rangle & \langle c_i^\dagger c_j^\dagger \rangle \\ 
        \langle c_i c_j \rangle & \langle c_i c_j^\dagger \rangle 
    \end{pmatrix} 
    =
    \begin{pmatrix} 
        C_{ij} & F_{ij} \\ 
        F_{ji}^* & \delta_{ij} - C_{ji} 
    \end{pmatrix},
\end{equation}
where $C$ and $F$ denote the normal and anomalous correlation matrix, respectively. To obtain the reduced density matrix $\rho_{\rm{A}}$, we can restrict the correlation matrix to subsystem A to obtain the block correlation matrix $G=\mathcal{G}|_{\text{A}}$. Consequently, the reduced density matrix of subsystem A is given by
\begin{equation}
    \label{E4}
    \rho_{\text{A}} =\det(I-G)\exp\left(\sum_{ij \in \text{A}}\left[\ln G(I-G)^{-1}\right]_{ij} \Psi_i^\dagger \Psi_j \right),
\end{equation}
where indices $i$ and $j$ belong only to subsystem A. By diagonalizing the block correlation matrix $G$, we obtain
\begin{equation}
    \label{E5}
    \rho_{\text{A}} =\exp\left(\sum_{k}\ln (1-\xi_k) + \sum_{n}\left[\ln \xi_n(1-\xi_n)^{-1} \right]\psi_n^\dagger \psi_n\right),
\end{equation}
where $\{\psi_n\}$ is a new basis obtained from $\{\Psi_n\}$ via an Bogoliubov transformation, and $\{\xi_n\}$ are the eigenvalues of block correlation matrix $G$, which are directly related to the entanglement spectrum~\cite{Zhou2022prr,chang2020prr,ye2016prb}. After this diagonalization, the reduced density matrix takes a diagonal form.

Thus, the entanglement entropy can be expressed in terms of the eigenvalues of the correlation matrix $\{ \xi_n \}$:
\begin{equation}
    \label{E6}
    S_{\rm{A}} = -\sum_n \left[ \xi_n \ln \xi_n + (1-\xi_n) \ln (1-\xi_n) \right].
\end{equation}

It is important to note that, due to particle-hole symmetry, the eigenvalues $\{ \xi_n \}$ appear in pairs of the form $\{ \xi_n, 1 - \xi_n \}$. As a result, when summing over the eigenvalues in Eq.~\eqref{E6}, they should first be arranged in ascending order, and the summation should be performed over only the first half of the spectrum.

\subsection{Global phase diagram from quantum entanglement}
\begin{figure}
    \centering
    \includegraphics[width=0.8\linewidth]{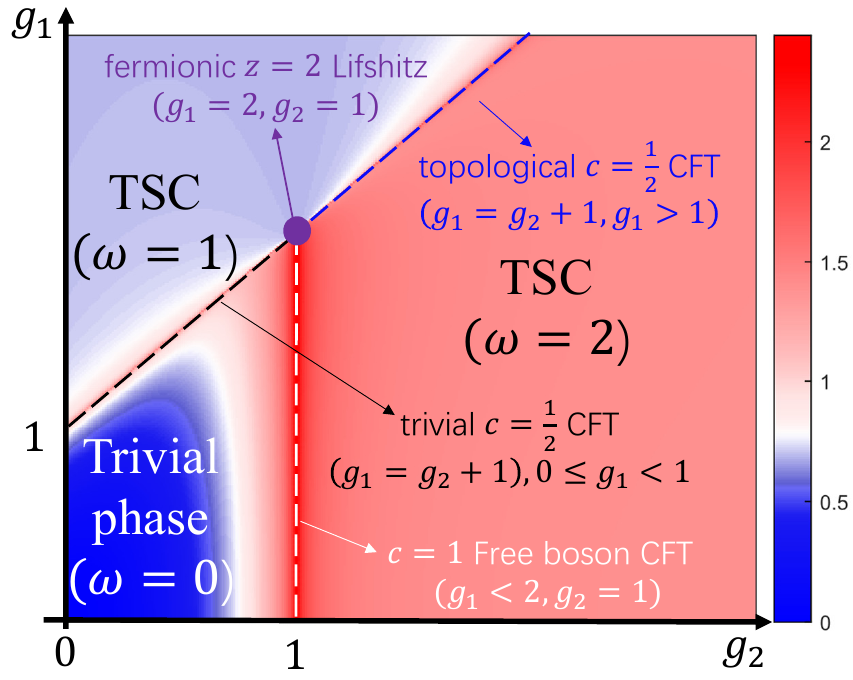}
    \caption{(Color Online) The global phase diagram of $ H = H_0 - g_1 H_1 + g_2 H_2 $ is shown as a function of the two control parameters $ g_1 $ and $ g_2 $. The TSC phases represent topological superconducting phases characterized by the winding number $ \omega $. The exact expressions for the three phase boundaries (depicted as dashed lines) can be determined analytically; see Appendix~\ref{sec:appA} for details. These transition lines are conformally invariant and exhibit distinct topological properties. The multicritical point (purple circle) is the only nonconformal critical point, described by the fermionic Lifshitz theory with a dynamical critical exponent $ z = 2 $. The colorbar indicates the value of the entanglement entropy. Simulations are performed with a system size of $ N = 400 $.}
    \label{fig2}
\end{figure}

Before delving into the details of the numerical results, let us first summarize the global quantum phase diagram of the model in Eq.~(\ref{E1}) from the entanglement perspective. The tuning parameters $(g_1, g_2)$ drive the system into different phases, including both topologically trivial and nontrivial phases characterized by different winding numbers, as illustrated in Fig.~\ref{fig2}.  The ground states of the fixed-point Hamiltonians $ H_{\alpha} $ with $ \alpha = 0,1,2 $ correspond to a trivial phase ($\alpha = 0$) and TSC phases with winding numbers $ \omega = 1 $ ($\alpha = 1$) and $ 2 $ ($\alpha = 2$), respectively. Away from the fixed-point limit, where competing interactions among $ g_0, g_1 $, and $ g_2 $ are present, these three topologically distinct phases extend over finite regions of the phase diagram, which is further confirmed by the entanglement entropy within each phase, as shown in the color plot in Fig.~\ref{fig2}. Specifically, we observe that within the phase regions, the entanglement entropy increases with the winding number. This is because a larger winding number indicates a larger number of fermionic edge modes under open boundary conditions, thereby contributing more entanglement in the TSC phases compared to the trivial phase.

\begin{figure*}
    \centering
    \includegraphics[width=1\linewidth]{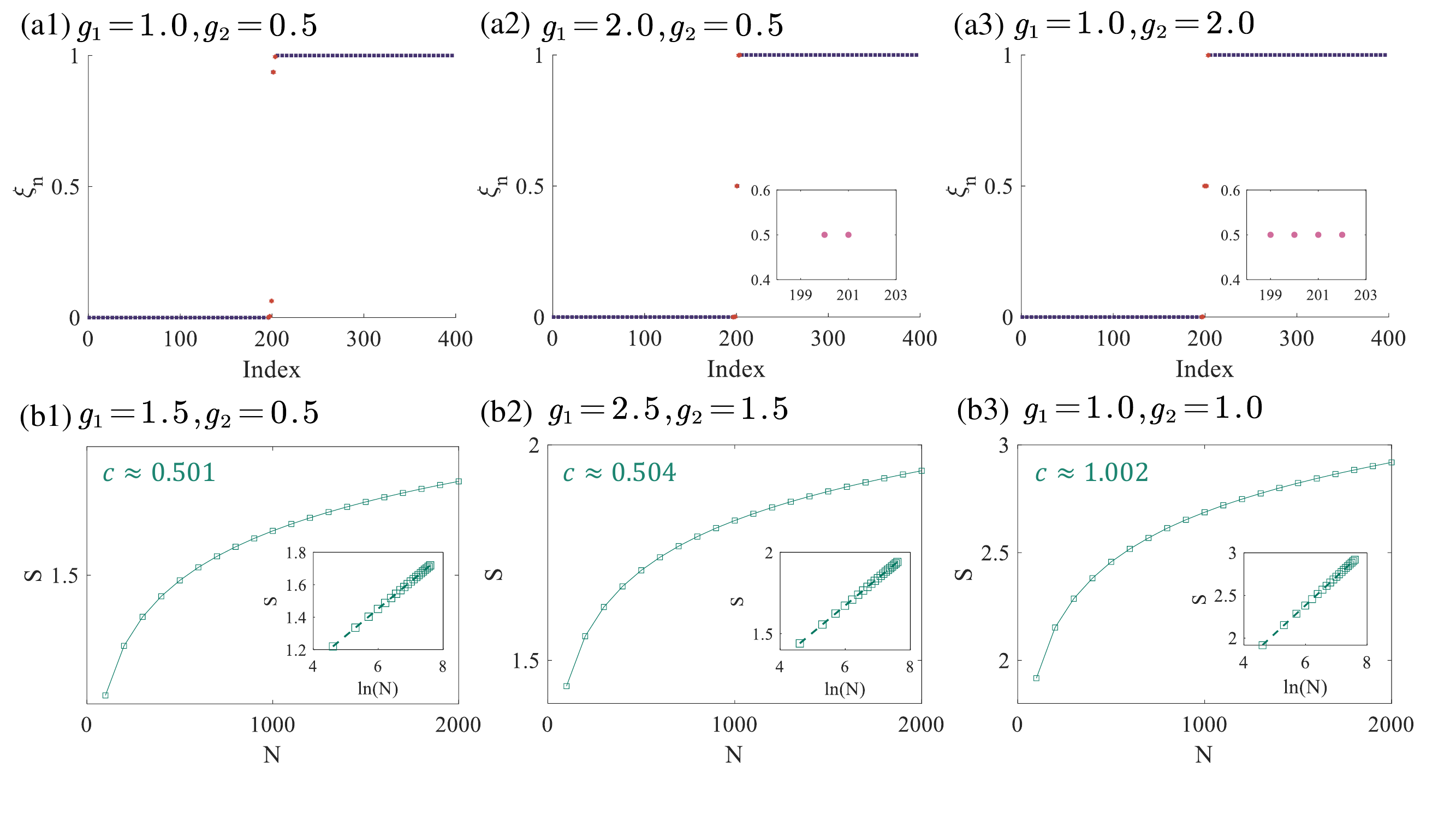}
    \caption{(Color Online) The entanglement spectrum $\{ \xi_n \}$ under periodic boundary conditions at representative points within the trivial (a1), $ \omega = 1 $ TSC (a2), and $ \omega = 2 $ TSC (a3) phases is shown. The insets of (a2) and (a3) display dangling degenerate edge modes, revealing the topological degeneracy in the entanglement spectrum, consistent with the Li-Haldane conjecture. Simulations are performed with a system size of $ N = 200 $. The insets show the degenerate edge modes by zooming in on isolated points in the spectrum. At selected points along the three conformal critical lines, we examine the scaling of the half-chain entanglement entropy $ S$ at the trivial (b1) and topologically nontrivial (b2) Majorana critical points, as well as at the Dirac universality class critical point (b3). The inset shows a log-log plot of $ S $, capturing the central charge of the underlying conformal critical point according to the relation $ S(N) \sim \frac{c}{3} \ln N + \text{const} $. The system size is chosen as $ N = 100, 200, \dots, 2000 $.}
    \label{fig3}
\end{figure*}

Moreover, three phase transitions exist between these topologically distinct quantum phases, as evidenced by the numerical results of entanglement entropy shown in Appendix~\ref{sec:fig7}. The transition points between the trivial and TSC phases, as well as between TSC phases with different winding numbers, exhibit topologically distinct Dirac cones~\cite{verresen2020topologyedgestatessurvive,Jones2019JSP}. Specifically, all the transition lines in the phase diagram can be analytically derived—see Appendix~\ref{sec:appA} for details.  

In summary, several key observations regarding the phase diagram are as follows:  

i) The conformal critical lines ($g_1 = g_2 + 1, 0 < g_2 < 1$) separating the trivial and $\omega = 1$ TSC phases belong to the topologically trivial Majorana universality class with central charge $c = 1/2$.  

ii) The conformal critical lines ($g_1 = g_2 + 1, g_2 > 1$) between the $\omega = 1$ and $\omega = 2$ TSC phases also belong to the $c = 1/2$ Majorana universality class but feature topologically protected edge modes under open boundary conditions. This class is now known as topological Majorana CFT.  

iii) The conformal critical lines ($g_1 = 1, 0 < g_2 < 2$) separating the trivial and $\omega = 2$ TSC phases belong to the Dirac universality class with central charge $c = 1$. 

iv) These three conformal critical lines intersect at a nonconformal multicritical point ($g_1 = 2, g_2 = 1$) with a dynamical exponent $z = 2$, known as the fermionic Lifshitz multicritical point.  

A systematic investigation of the entanglement properties of these nontrivial critical states is the primary focus of this paper, which we will gradually reveal in the following sections.

\subsection{Entanglement spectrum in gapped fermionic topological phase}
The nontrivial topological degeneracy can be encoded in the bulk wavefunction without requiring any boundaries. A well-known principle of bulk-boundary correspondence is the Li-Haldane conjecture~\cite{Li2008prl}, which states that the low-lying entanglement spectrum of a bulk wavefunction can be exactly reconstructed from the energy spectrum under open boundary conditions. This conjecture implies that although topological and trivial phases were traditionally considered indistinguishable based on bulk local observables, they can, in fact, be distinguished through their bulk entanglement spectrum. More broadly, the Li-Haldane conjecture was originally proposed to investigate the topological properties of fractional quantum Hall states. It was later generalized to both non-interacting and interacting gapped SPT phases~\cite{Pollmann2010prb,Turner2011prb,Qi2012prl,Fidkowski2010prl,Hsieh2014prl,Cho_2017}, and more recently, to interacting bosonic symmetry-enriched QCPs~\cite{yu2024prl,zhang2024pra}.

This work aims to employ quantum entanglement diagnostics to probe topological degeneracy in the Majorana $\alpha$ chain described by Eq.~(\ref{E1}). For completeness, we first focus on fermionic gapped topological phases, which have been extensively studied in the literature~\cite{Wen2017rmp}. We select one representative point from each region of the phase diagram and compute the bulk entanglement spectrum $\{ \xi_n \}$, as shown in Fig.~\ref{fig3} (a1-a3). For $g_1=1.0, g_2=0.5$, the bulk entanglement spectrum does not exhibit any degeneracy (Fig.~\ref{fig3}(a1)), indicating that this region corresponds to a topologically trivial phase. In contrast, for $g_1=2.0, g_2=0.5$ and $g_1=1.0, g_2=2.0$, the bulk entanglement spectrum exhibits two-fold and four-fold degeneracy, respectively, implying that these regions correspond to topologically nontrivial phases with winding numbers $\omega=1$ and $\omega=2$ (see the insets of Fig.~\ref{fig3}(a2-a3)). Furthermore, as confirmed in Appendix~\ref{sec:Appendix_B}, these topological edge modes are also reflected in the energy spectrum under open boundary conditions, providing numerical verification of the Li-Haldane bulk-boundary correspondence in fermionic SPT states. Additional numerical evidence at other points in the phase diagram can be found in Appendix~\ref{sec:fig8}.

\subsection{Entanglement entropy and spectrum in fermionic symmetry-enriched QCPs}

\subsubsection{Entanglement entropy}
After exploring the entanglement properties of all quantum phases, we now shift our focus to the more intriguing QCPs between these phases from the perspective of entanglement. To explore the universality along the transition lines, we first examine the half-chain entanglement entropy $S$ at three critical points in the phase diagram. As shown in Fig.~\ref{fig3} (b1-b3), we select three representative points along the transition lines and compute the scaling behavior of $S$ for different system sizes $N$. For $g_1 = 1.5, g_2 = 0.5$ and $g_1 = 2.5, g_2 = 1.5$, which correspond to transition points between a TSC with winding number $\omega = 1$ and a trivial phase (Fig.~\ref{fig3} (b1)) or between a TSC with $\omega = 1$ and another with $\omega = 2$ (Fig.~\ref{fig3} (b2)), respectively, the numerical results unambiguously demonstrate that the entanglement entropy at these critical points follows a logarithmic scaling (see the insets of Fig.~\ref{fig3} (b1-b3)). This confirms that the transitions are described by a conformally invariant field theory with central charge $c = 1/2$, placing them in the Majorana universality class~\cite{francesco2012conformal,ginsparg1988appliedconformalfieldtheory}. Conversely, for the transition point at $g_1 = 1.0, g_2 = 1.0$, which separates the trivial phase from a TSC phase with winding number $\omega = 2$ (Fig.~\ref{fig3} (b3)), the entanglement entropy at criticality also follows a logarithmic scaling but with a central charge of $c = 1$, indicating that this transition belongs to the Dirac universality class~\cite{verresen2018prl,yu2024prb}. Additional numerical evidence supporting these findings for other points along the transition lines is provided in Appendix~\ref{sec:fig7}. Interestingly, for quantum critical lines that are topologically distinct—such as the two $c = 1/2$ Majorana critical lines mentioned above—entanglement entropy alone fails to distinguish the underlying nontrivial topology. This limitation can be overcome by analyzing the bulk entanglement spectrum, which we discuss in the next subsection. Finally, the entanglement entropy at the Lifshitz multicritical point ($g_1 = 2, g_2 = 1$) exhibits anomalous scaling behavior due to the absence of conformal invariance, which has recently attracted great attention~\cite{Wangke2022scipost,Wangke2023prb}.

\begin{figure}
    \centering
    \includegraphics[width=1\linewidth]{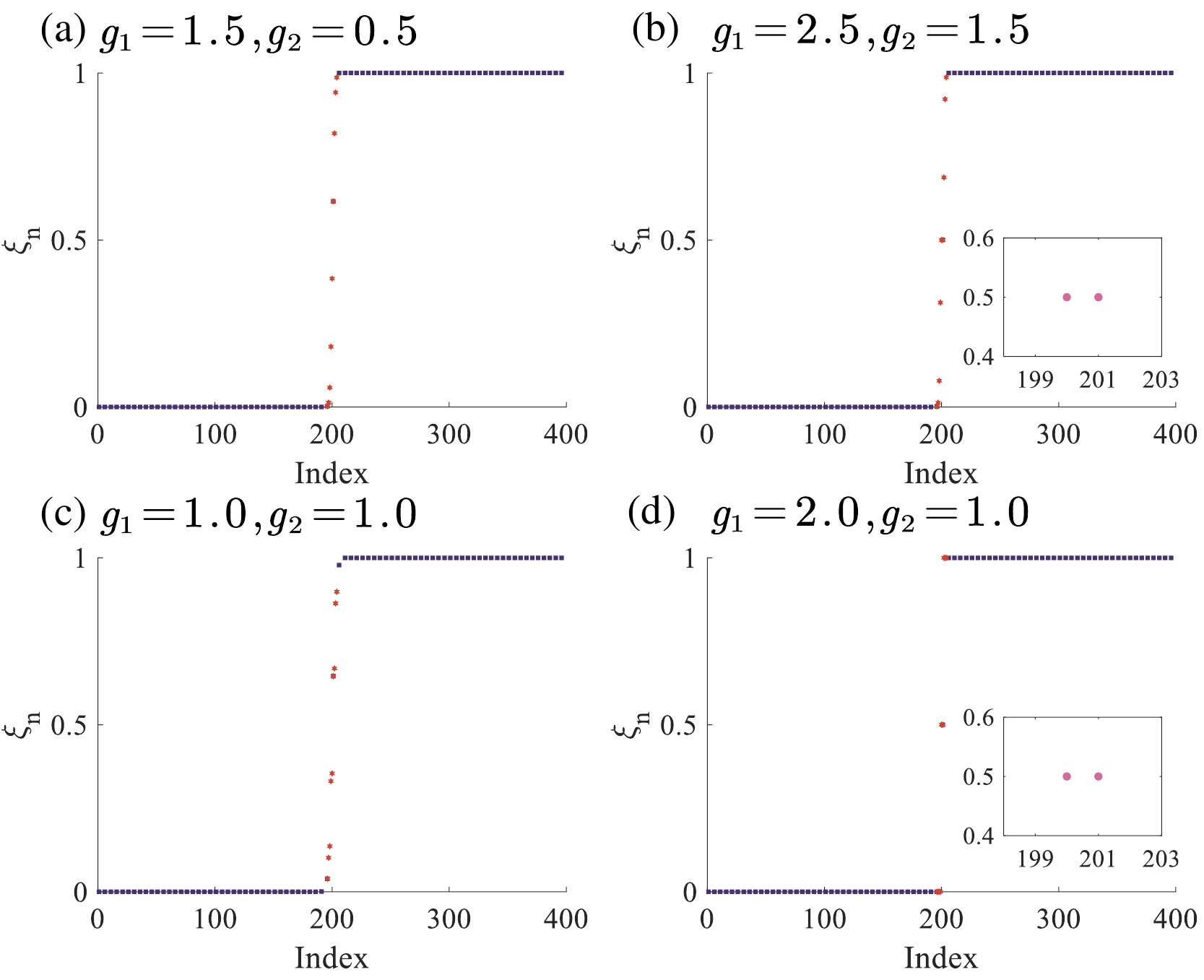}
    \caption{(Color Online) The entanglement spectrum at four different critical points is shown. (a) The conformal critical point between the trivial and $ \omega = 1 $ TSC phases. (b) The conformal critical point between the $ \omega = 1 $ and $ \omega = 2 $ TSC phases. (c) The conformal critical point between the trivial and $ \omega = 2 $ TSC phases. (d) The nonconformal multicritical point at the intersection of the three critical lines. The insets show the degenerate edge modes by zooming in on isolated points in the spectrum. Simulations are performed with a system size of $ N = 200 $.}
    \label{fig4}
\end{figure}

\subsubsection{Entanglement spectrum}
To reveal the nontrivial topology along the quantum critical lines, we examine the bulk entanglement spectrum to identify possible topological degeneracies based on the Li-Haldane bulk-boundary correspondence. Specifically, we first focus on the entanglement spectrum at the transition point between trivial and TSC phases with winding numbers $\omega = 1$ and $2$, respectively. As illustrated in Fig.~\ref{fig4}(a) and (c), we observe that the bulk entanglement spectrum does not exhibit any degeneracy, which is consistent with the features of the energy spectrum under open boundary conditions, as shown in Appendix~\ref{sec:Appendix_B}. These results imply that both critical points between trivial and TSC phases with different winding numbers belong to a topologically trivial universality class, in agreement with previous theoretical arguments~\cite{verresen2018prl,yu2024prb}. Conversely, the entanglement spectra at the remaining critical and multicritical points, shown in Fig.~\ref{fig4}(b) and (d), exhibit nontrivial topological degeneracy. More precisely, our numerical results unambiguously demonstrate that the entanglement spectrum of the conformal critical point between TSC phases with winding numbers $\omega = 1$ and $2$—and, more surprisingly, the nonconformal Lifshitz multicritical point—both display two-fold topological degeneracy in the bulk entanglement spectrum. This degeneracy is also consistent with the energy spectrum under open boundary conditions, further supporting the predictions of the Li-Haldane bulk-boundary correspondence. Additional numerical evidence for the entanglement spectra at other points along the critical line can be found in Appendix~\ref{sec:fig8}. Consequently, we demonstrate that the bulk entanglement spectrum serves as an effective diagnostic tool for distinguishing fermionic symmetry-enriched QCPs in non-interacting systems.

Finally, we would like to make one additional remark: as in the case of gapped SPT phases, the classification of bosonic and fermionic SPT phases generally follows distinct theoretical frameworks~\cite{Chen2013prb,Chen2012Science,Bi2015PRB,senthil2015symmetry,Lan2017PRB,Cheng2018PRB}. Therefore, it is essential to investigate the topological properties of bosonic and fermionic symmetry-enriched QCPs separately. Our previous work~\cite{yu2024prl} focused on the entanglement spectrum of bosonic symmetry-enriched QCPs in one-dimensional, strongly interacting quantum spin chains. Extending the notion of gapless topology to higher-dimensional, strongly interacting systems presents considerable analytical and numerical challenges due to the lack of well-established theoretical frameworks and efficient many-body computational algorithm. In contrast, free fermion critical systems do not suffer from these limitations, making them a promising platform for studying higher-dimensional symmetry-enriched QCPs. In Appendix~\ref{appE}, we present additional numerical simulations and a brief discussion of two-dimensional fermionic symmetry-enriched QCPs from the entanglement perspective. A more systematic study of higher-dimensional symmetry-enriched QCPs, particularly in interacting systems, is considerably more involved and beyond the scope of the present work, but we regard it as an important direction for future research.

\section{CONCLUSION AND OUTLOOK}
\label{sec:con}
To summarize, using the critical Majorana $\alpha$-chain with time-reversal and particle-hole symmetries as an illustrative example, we have explored quantum entanglement properties and established a bulk-boundary correspondence reflected in the bulk entanglement spectrum for both gapped and gapless fermionic topological states. Specifically, employing entanglement entropy and spectrum as diagnostic tools, we obtain a global phase diagram for the Hamiltonian, which interpolates between different fixed-point Hamiltonians for $H_0$, $H_1$, and $H_2$. The competition between tuning parameters gives rise to trivial and topologically nontrivial gapped phases with winding numbers $\omega = 1$ and $2$, as confirmed numerically through the topological degeneracy in the bulk entanglement spectrum, in accordance with the Li-Haldane conjecture. More importantly, the transition lines in the phase diagram can be analytically determined due to the solvability of the free fermion model. By analyzing the scaling of entanglement entropy, we find that both transitions between the $\omega = 1$ TSC phase and either the trivial or $\omega = 2$ TSC phase belong to the Majorana universality class with central charge $c = 1/2$, whereas the transition between the trivial and $\omega = 2$ TSC phase belongs to the Dirac universality class with $c = 1$. Furthermore, the topological distinctions between these three critical lines can be revealed through the nontrivial degeneracy of the bulk entanglement spectrum, providing an effective diagnostic for probing nontrivial topology based solely on bulk quantities and generalizing the Li-Haldane bulk-boundary correspondence to fermionic symmetry-enriched QCPs. Additionally, we examine the topological properties at the intersection of the three transition lines and identify a topological Lifshitz multicritical point characterized by two-fold degeneracy in the entanglement spectrum. Future directions of interest include extending the bulk-boundary correspondence in entanglement spectra to higher dimensions and developing a comprehensive theory of quantum entanglement in symmetry-enriched QCPs, at least for free fermion systems. Our work sheds new light on the underlying mechanisms of quantum entanglement in gapless topological phases of matter.

\begin{acknowledgments}
X.-J.Yu thank Shao-Kai Jian and Sheng Yang for collaboration on related projects. We thank Zhi-Kang Lin for helpful discussions. X.-J. Yu was supported by the National Natural Science Foundation of China (Grant No.12405034) and a start-up grant from Fuzhou University. H.Q. Lin acknowledges financial support from 
NSFC12088101 and MOST 2022YFA1402701. 
\end{acknowledgments}

\bibliography{main}

\newpage
\onecolumngrid

\appendix

\section{EXACT SOLUTION OF MAJORANA $\alpha$ CHAIN}
\label{sec:appA}
In this section, we provide a detailed analytical solution for the Hamiltonian $H=H_0-g_1H_1+g_2H_2$. We begin by substituting the relations $\gamma_n=c_n^\dagger+c_n$ and $\tilde{\gamma}_n=i(c_n^\dagger-c_n)$ into the complex fermion $c$ and $c^\dagger$, yielding
\begin{equation}
    \label{A1}
    H=\sum_n \left[ (1-2c_n^\dagger c_n) +g_1(c_n^\dagger c_{n+1}+c_n^\dagger c_{n+1}^\dagger +h.c.)-g_2(c_n^\dagger c_{n+2}+c_n^\dagger c_{n+2}^\dagger +h.c.)\right],
\end{equation}
here $h.c.$ means hermitian conjugate. Then we can derive further result by using Fourier transformation $c_{n}=\frac{1}{\sqrt{N}}\sum_k \text{e}^{-ikn}c_k$, where $k=\pm \frac{(2n-1)\pi}{2},n=1,2...N/2$. Thus the Hamiltonian $H$ in k-space is 
\begin{equation}
    \label{A2}
    H=\sum_k \left[ iy_k \left(c_kc_{-k}+c_k^\dagger c_{-k}^\dagger \right) +z_k\left(c_k^\dagger c_{k}+c_{-k}^\dagger c_{-k} \right) \right],
\end{equation}
here, $y_k=g_1\sin k -g_2 \sin 2k$, and $z_k=g_1\cos k -g_2\cos 2k -1$. To diagonalize the Hamiltonian, we introduce the Bogoliubov transformation: 
\begin{align}
    \label{A3}
    & a_k = \cos(\frac{\theta_k}{2}) c_k - i\sin(\frac{\theta_k}{2}) c_{-k}^\dagger\\
    & a_k^\dagger = \cos(\frac{\theta_k}{2}) c_k^\dagger + i\sin(\frac{\theta_k}{2}) c_{-k},
\end{align}
with $\tan \theta_k=-\frac{y_k}{z_k}$. The Hamiltonian can be diagonalized as 
\begin{equation}
    \label{A5}
    H=\sum_{k>0} \varepsilon_k \left( a_k^\dagger a_k-\frac{1}{2} \right),
\end{equation}
where the quasiparticle dispersion is given by $\varepsilon_k=4\sqrt{y_k^2+z_k^2}$. A phase transition occurs when the energy gap closes, i.e., when $\varepsilon_k=0$, which requires both $y_k =0 $ and $z_k = 0$. At these gapless points, low-energy excitations emerge, leading to singularities in the free energy and other thermodynamic quantities. Therefore, the condition $y_k=z_k=0$ marks the presence of a gapless excitation, indicating a quantum phase transition.

Moreover, the phase diagram exhibits three quantum phases and the phase transitions between them, as discussed in Sec.~\ref{sec:model} of the main text. We now derive the exact expression for the phase boundary lines in the phase diagram in detail.

We express the equations $ y_k = z_k = 0 $ as  
\begin{equation}
    \label{A6}
     g_1\sin k - g_2\sin 2k = \sin k (g_1 - 2g_2\cos k) = 0,
\end{equation}  
and  
\begin{equation}
    \label{A7} 
     g_1\cos k - g_2\cos 2k = 1.
\end{equation}  

From Eq.~\eqref{A6}, there are two possible solutions: $ k = 0 $ and $ \cos k = g_1/(2g_2) $. First, substituting $ k = 0 $ into Eq.~\eqref{A7}, we immediately obtain $ g_1 = g_2 + 1 $. This result corresponds to the black and blue dashed lines in Fig.~\ref{fig2}. On the other hand, substituting $ \cos k = g_1/(2g_2) $ into Eq.~\eqref{A7} and imposing the constraint $ 0 < \cos k < 1 $, we find that $ 0< \color{black} g_1 < 2 $ and $ g_2 = 1 $. This result corresponds to the white dashed line in Fig.~\ref{fig2}. Additionally, there is a critical point at $ (g_1, g_2) = (2,1) $, where the conditions $ g_1 = g_2 + 1 $ and $ g_2 = 1 $ are simultaneously satisfied. This point is represented by the purple solid circle in Fig.~\ref{fig2}. However, we cannot conclusively determine the universality class of these three phase transition lines or whether they exhibit nontrivial topological properties. Addressing this question is a central focus of our work and underscores the importance of analyzing the quantum entanglement properties of these systems.

\section{NUMERICAL RESULTS FOR ENERGY SPECTRUM UNDER OBCs}
\label{sec:Appendix_B}

\begin{figure*}
    \centering
    \includegraphics[width=0.6\linewidth]{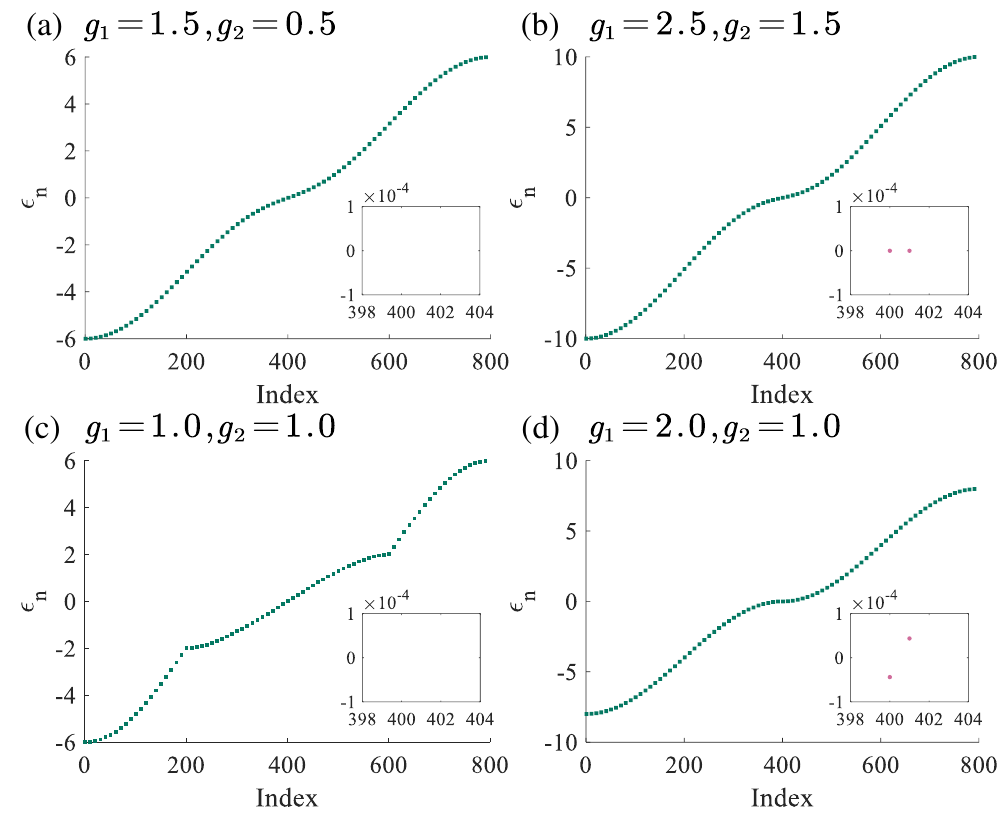}
    \caption{The open-boundary energy spectrum is shown for the critical points: (a) $ (g_1, g_2) = (1.5, 0.5) $, (b) $ (g_1, g_2) = (2.5, 1.5) $, (c) $ (g_1, g_2) = (1.0, 1.0) $, and (d) $ (g_1, g_2) = (2.0, 1.0) $. The insets show the degenerate edge modes by zooming in on isolated points in the spectrum. Simulations are performed for a system size of $ N = 400 $, displaying data points at intervals of $10$ sites, except for the edge states, which are highlighted as purple circle points.}
        \label{fig5}
\end{figure*}

\begin{figure*}
    \centering
    \includegraphics[width=1\linewidth]{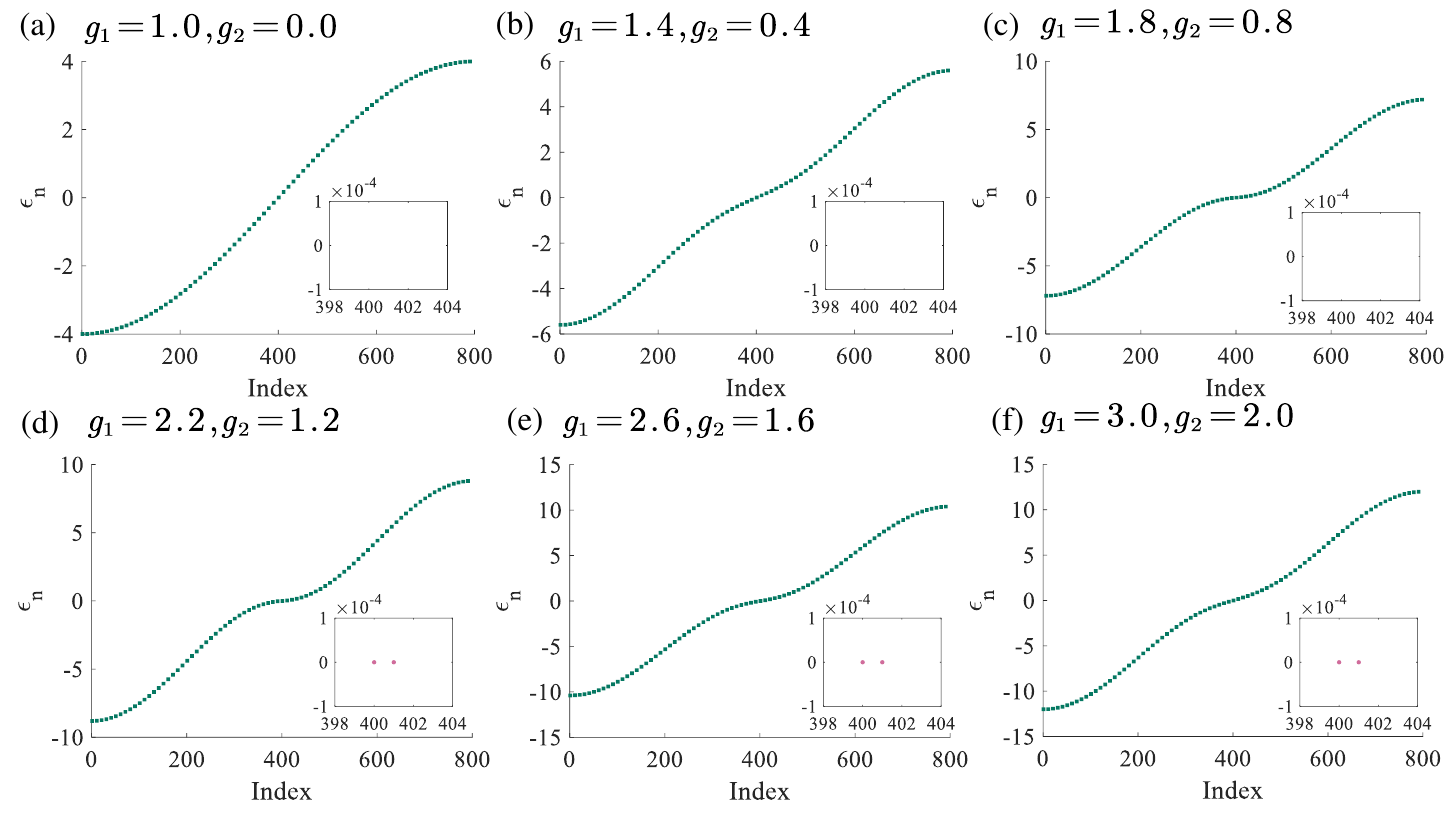}
    \caption{The open-boundary energy spectrum is shown for additional critical points along three transition lines: (a)(b)(c) along the conformal critical line $ (g_1 = g_2 + 1,\, 0 < g_2 < 1) $, and (d)(e)(f) along the conformal critical line $ (g_1 = g_2 + 1,\, g_2 > 1) $. The insets show the degenerate edge modes by zooming in on isolated points in the spectrum. Simulations are performed for a system size of $ N = 400 $, displaying data points at intervals of $10$ sites, except for the edge states, which are highlighted as purple circle points.}
    \label{fig6}
\end{figure*}

In this section, we provide additional numerical data to display the energy spectrum under open boundary conditions (OBCs).

The Hamiltonian can be diagonalized as $H=\sum_{n=1}^{N} \epsilon_{n}\eta^{\dagger}_{n}\eta_{n}$ through a canonical Bogoliubov transformation by introducing the fermionic operators $\eta_{n}$ and $\eta^{\dagger}_{n}$,

\begin{equation}
\eta_{n}=\sum^{N}_{i}(u^{*}_{n,i}c_{i}+v_{n,i}c^{\dagger}_{i}),~\eta^{\dagger}_{n}=\sum^{N}_{i}(u_{n,i}c^{\dagger}_{i}+v^{*}_{n,i}c_{i}),
\end{equation}
where $u_{n,i}$ and $v_{n,i}$ denote the two components of the wave function at site $j$, $n$ is the energy band index, and $\epsilon_{n}$ represents the eigenstate energy. The Schr\"{o}dinger equation $H\ket{\Psi_{n}}=\epsilon_{n}\ket{\Psi_{n}}$ can be written as

\begin{equation}
\label{E10}
\begin{pmatrix}
A & B \\
-B^{*} & -A^{T}
\end{pmatrix}
\begin{pmatrix}
u_{n,i} \\
v_{n,i}^{*}
\end{pmatrix}=
\epsilon_{n}
\begin{pmatrix}
u_{n,i} \\
v_{n,i}^{*}
\end{pmatrix},
\end{equation}
where $A(B)$ is a $N\times N$ symmetric (antisymmetric) matrix, the zero energy state probability distributions can be computed as $\lvert \Psi_{n,i}\rvert^{2}=\lvert u_{n,i}\rvert^{2}+\lvert v_{n,i}\rvert^{2}$. 

To establish the Li-Haldane bulk-boundary correspondence~\cite{Li2008prl}, which states that the low-lying structure of the bulk entanglement spectrum is equivalent to the energy spectrum under open boundary conditions, we select the same parameter points as those used for the entanglement spectrum calculations in the main text: (a) $ (g_1, g_2) = (1.5, 0.5) $, (b) $ (g_1, g_2) = (2.5, 1.5) $, (c) $ (g_1, g_2) = (1.0, 1.0) $, and (d) $ (g_1, g_2) = (2.0, 1.0) $. The numerical results for the energy spectrum as a function of the state index are shown in Fig.~\ref{fig5}. Comparing these results with Fig.~\ref{fig4} in the main text, we observe that at the critical points lying on the three different conformal critical lines (Fig.~\ref{fig4}(a)(b)(c) and Fig.~\ref{fig5}(a)(b)(c)), both spectra exhibit the same two-fold topological degeneracy, consistent with the Li-Haldane conjecture, even in fermionic symmetry-enriched QCPs. However, at the nonconformal Lifshitz multicritical point, the Li-Haldane bulk-boundary correspondence appears to break down, as indicated by the absence of exact degeneracy in the open-boundary energy spectrum, despite the presence of double degeneracy in the bulk entanglement spectrum discussed in the main text. Consequently, we termed the multicritical point as topologically nontrivial in the sense that it exhibits nontrivial degeneracy in the bulk energy spectrum. A more detailed and systematic study of the relationship between the energy spectrum and the entanglement spectrum is left for future work.

Additionally, we present numerical results for the energy spectrum under open boundary conditions for additional points along the three transition lines, as shown in Fig.~\ref{fig6}. Combining these results with the bulk energy spectrum at shown in Fig.~\ref{fig8}, our results unambiguously demonstrate that the three conformal critical lines are topologically distinct and satisfy the Li-Haldane bulk-boundary correspondence.

\section{ADDITIONAL DATA FOR ENTANGLEMENT ENTROPY}
\label{sec:fig7}

\begin{figure*}
    \centering
    \includegraphics[width=1\linewidth]{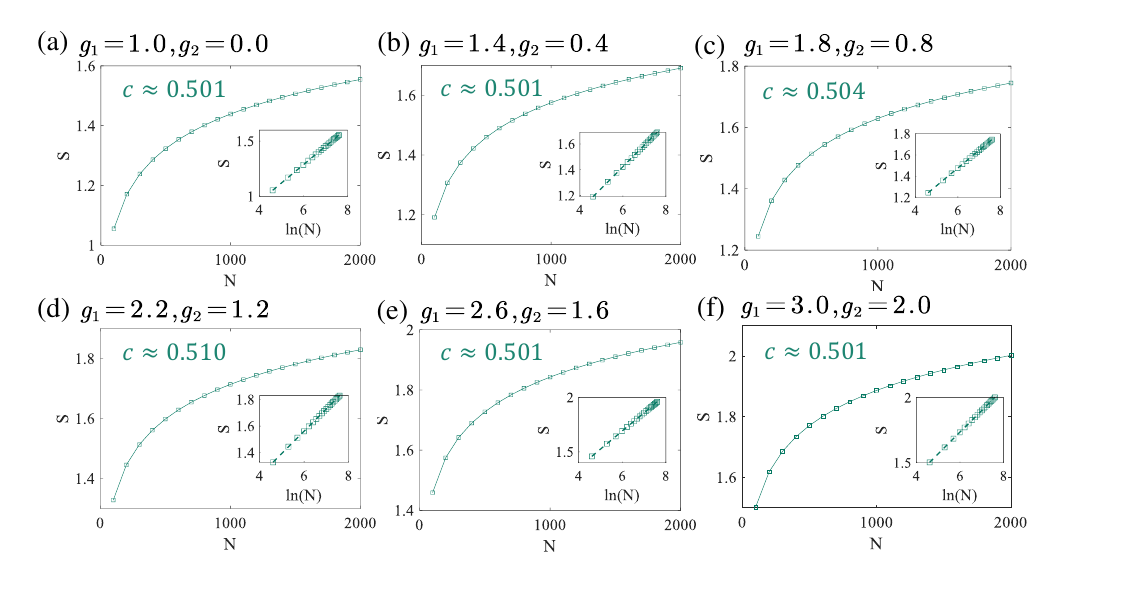}
    \caption{The entanglement entropy is shown for additional critical points along three transition lines: (a)(b)(c) along the conformal critical line $ (g_1 = g_2 + 1,\, 0 < g_2 < 1) $ and (d)(e)(f) along the conformal critical line $ (g_1 = g_2 + 1,\, g_2 > 1) $. The insets display the finite-size scaling of the entanglement entropy following the scaling relation $ S \sim \frac{c}{3} \ln N + S_0 $. The system size is chosen as $ N = 100, 200, \dots, 2000 $.}
        \label{fig7}
\end{figure*}

In this section, we provide additional numerical data for the half-chain entanglement entropy $ S$. Similar to the analysis in the main text, we present the finite-size scaling of $ S$ at additional critical points along the three transition lines in Fig.~\ref{fig7}: (a) $ (g_1, g_2) = (1.0, 0.0) $, (b) $ (g_1, g_2) = (1.4, 0.4) $, and (c) $ (g_1, g_2) = (1.8, 0.8) $; (d) $ (g_1, g_2) = (2.2, 1.2) $, (e) $ (g_1, g_2) = (2.6, 1.6) $, and (f) $ (g_1, g_2) = (3.0, 2.0) $. The numerical results confirm that all three critical lines exhibit conformal invariance, as evidenced by the logarithmic scaling of the entanglement entropy. More importantly, the extracted central charge from the scaling relation $ S \sim \frac{c}{3} \ln N + S_0 $ aligns with the expected Majorana or Dirac universality class, corresponding to $ c = 1/2 $ or $ c = 1 $, respectively.

\section{ADDITIONAL DATA FOR ENTANGLEMENT SPECTRUM}
\label{sec:fig8}

\begin{figure*}
    \centering
    \includegraphics[width=1\linewidth]{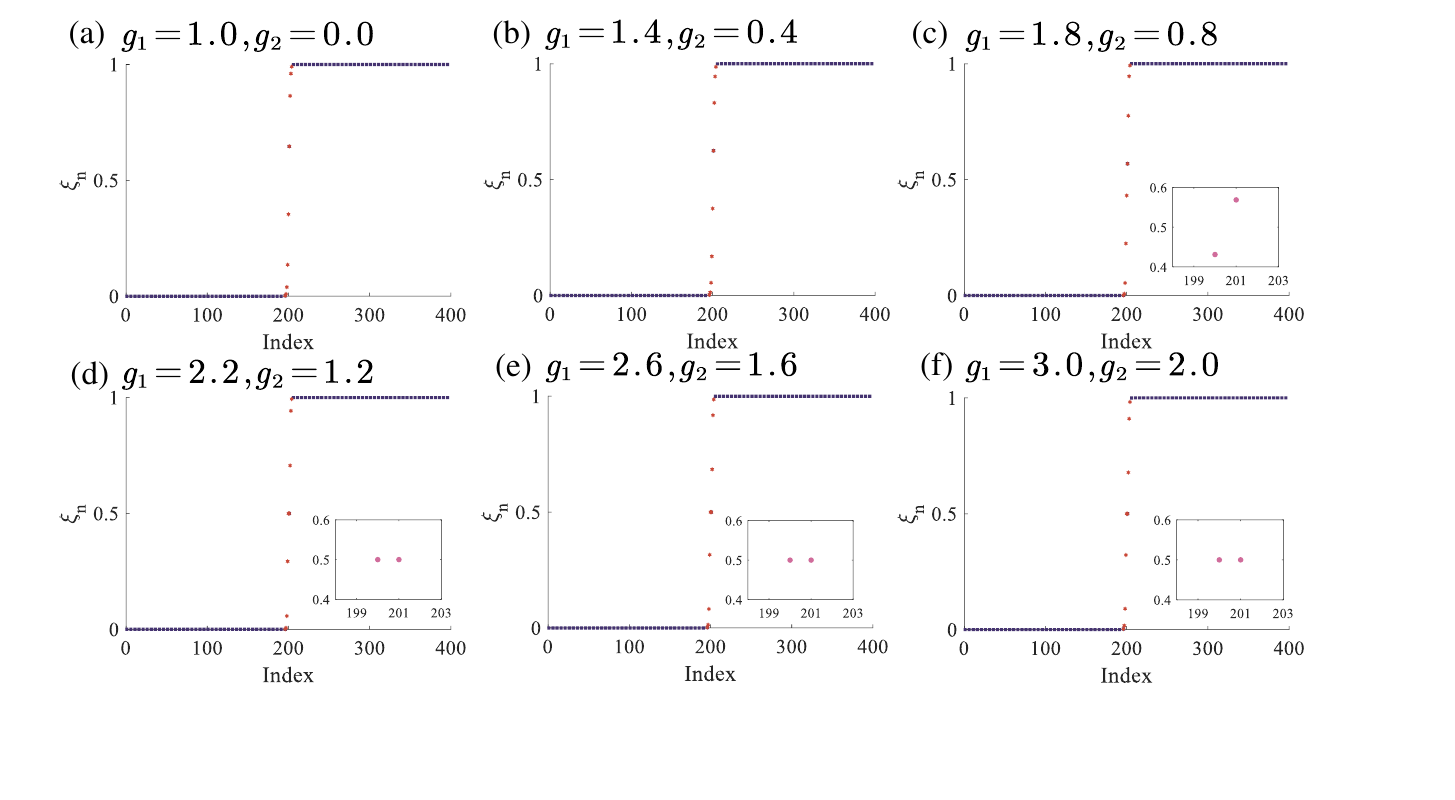}
    \caption{The bulk entanglement spectrum is shown for additional critical points along three transition lines: (a)(b)(c) along the conformal critical line $(g_1 = g_2 + 1,\, 0 < g_2 < 1) $ and (d)(e)(f) along the conformal critical line $ (g_1 = g_2 + 1,\, g_2 > 1) $. The insets show the degenerate edge modes by zooming in on isolated points in the spectrum. Simulations are performed for a system size of $ N = 200 $.}
    \label{fig8}
\end{figure*}

In this section, we provide additional numerical data on the bulk entanglement spectrum under periodic boundary conditions. Similar to the main text, the entanglement spectrum is directly related to the eigenvalues $ \{ \xi_n \} $ of the block correlation function, defined as $ G = \mathcal{G}|_\text{A} $. Specifically, we numerically compute the entanglement spectrum and present $ \xi_n $ as a function of the state index for simplicity. Additional points along the three transition lines are shown in Fig.~\ref{fig8}: (a) $ (g_1,g_2) = (1.0,0.0) $, (b) $ (g_1,g_2) = (1.4,0.4) $, (c) $ (g_1,g_2) = (1.8,0.8) $, (d) $ (g_1,g_2) = (2.2,1.2) $, (e) $ (g_1,g_2) = (2.6,1.6) $, and (f) $ (g_1,g_2) = (3.0,2.0) $. The numerical results confirm the emergence of fermionically topological distinct quantum critical lines in the phase diagram and provide further evidence that the entanglement spectrum serves as an effective diagnostic for detecting nontrivial topology at criticality.

\section{QUANTUM ENTANGLEMENT IN TOPOLOGICALLY DISTINCT CHERN CRITICALITY IN TWO DIMENSION}
\label{appE}
In this section,we provide additional numerical evidence to demonstrate nontrivial gapless topology in higher dimension can be identified from entanglement point of view. We consider a two dimensional fermionic lattice model for Chern insulator transition~\cite{verresen2020topologyedgestatessurvive}, which can be expressed in the momentum space under PBC:
\begin{equation}
H_\alpha = \sum_{\bm k} \left( c_{\bm k,A}^\dagger, c_{\bm k,B}^\dagger \right) \mathcal H(\bm k) \left( \begin{array}{c}
c_{\bm k,A} \\
c_{\bm k,B}
\end{array} \right) \qquad \textrm{with } \mathcal H ( \bm k ) = (\sin(k_x)\sigma_x-\sin(k_y))\sigma_y + (-\Delta-\frac{\cos(k_x)}{2}-\frac{\cos(k_y)}{2}-t\cos(k_x+k_y))\sigma_z, \label{eq:Halpha}
\end{equation}

\begin{figure*}
    \centering
    \includegraphics[width=1\linewidth]{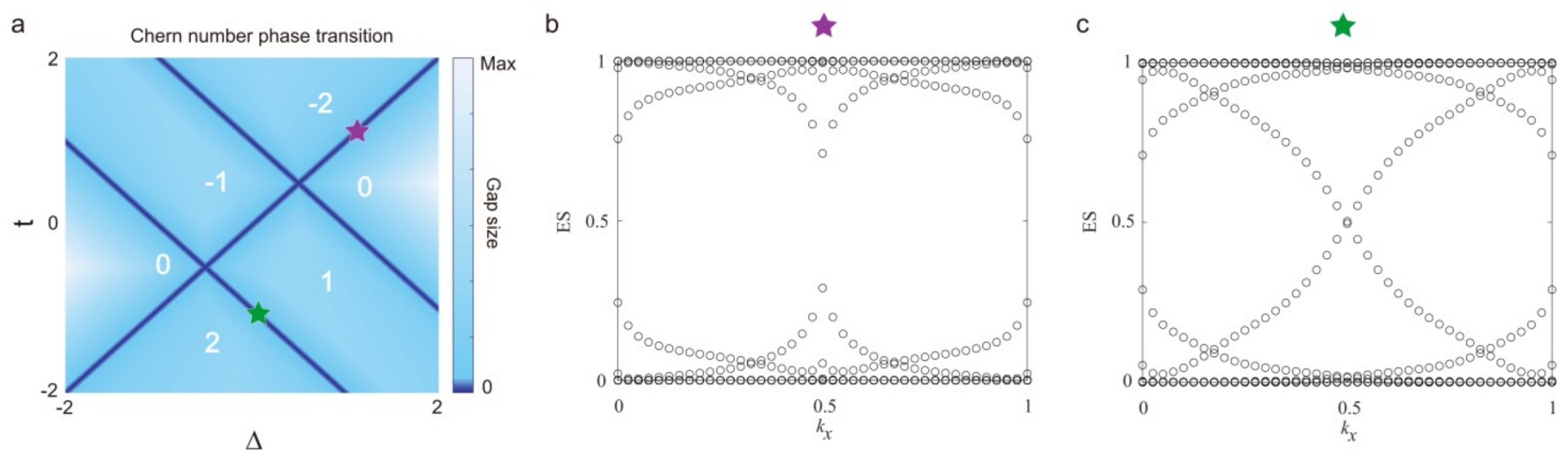}
    \caption{(a) Phase diagram of a two dimensional fermionic lattice model Eq.(~\eqref{eq:Halpha}) (b) Bulk entanglement spectrum at the critical point between phases with Chern numbers 0 and -2 ($t=1.0$, $\Delta=1.0$), corresponding to the purple star in the panel (a). (c) Bulk entanglement spectrum at the critical point between phases with Chern numbers 1 and 2 ($t=-1.0$, $\Delta=0.0$ ), corresponding to the green star in the panel (a). All simulations are performed on systems with $N=40\times 40$ unit cells, with the entanglement computed for a subsystem of $40 \times 10$ unit cells.}
    \label{fig9}
\end{figure*}

where $t$ and $\Delta$ serve as tuning parameters, with the resulting phase diagram displayed in Fig.~\ref{fig9} (a). We focus on the direct transitions between $\mathcal C =0 \leftrightarrow \mathcal C=-2$ and $\mathcal C=1 \leftrightarrow \mathcal C=2$ are topologically distinct Dirac cones, respectively. To investigate the topological properties at criticality, we compute bulk entanglement spectrum at two topologically distinct critical points, as shown in Fig.~\ref{fig9} (b) and (c). Specifically, the entanglement spectrum is plotted as a function of $k_x$, indicating that the system has open boundary conditions along the $y$-direction and periodic boundary conditions along the $x$-direction. The entanglement spectrum at the critical point $(t, \Delta) = (-1.0, 0.0)$ unambiguity displays chiral edge states (midgap states in the spectra), which are absent at $(t, \Delta) = (1, 1)$. This contrast not only reinforces the utility of the entanglement spectrum as a diagnostic tool for distinguishing fermionic symmetry-enriched QCPs in higher dimensions but also highlights a promising route for systematically studying higher-dimensional fermionic symmetry-enriched QCPs without resorting to large-scale many-body computations.

\end{document}